\documentclass[10pt,twocolumn]{IEEEtran}

\usepackage{subfigure}
\usepackage{amsmath}
\usepackage{amsfonts}
\usepackage{mathrsfs}
\usepackage[dvips]{graphicx}

\usepackage{epsfig}
\usepackage{color}
\usepackage{graphicx}

\begin{document}

\newtheorem{thm}{Theorem}
\newtheorem{prop}{Proposition}
\newtheorem{lem}{Lemma}

\title{Enhancement of Secrecy of Block Ciphered Systems by Deliberate Noise}

\author{Yahya S. Khiabani, Shuangqing Wei, Jian Yuan and  Jian Wang}
\maketitle

\footnotetext[1]{Y. Khiabani and S. Wei are with the Department of
Electrical and Computer Engineering, Louisiana State University, Baton
Rouge, LA 70803, USA (Email: ysowti1@tigers.lsu.edu; swei@lsu.edu). Their
work is supported in part by the Board of Regents of Louisiana under
contract LEQSF(2009-11)-RD-B-03.  J. Yuan and J. Wang are with the
Department of Electronic Engineering, Tsinghua University, Beijing,
P. R. China, 100084. (E-mail: {jyuan, jian-wang}@tsinghua.edu.cn)}

\begin{abstract} 

This paper considers the problem of end-end security enhancement by
resorting to deliberate noise injected in ciphertexts. The main goal is
to generate a degraded wiretap channel in application layer over which
Wyner-type secrecy encoding is invoked to deliver additional secure
information. More specifically,  we study secrecy enhancement of DES
block cipher working in cipher feedback model (CFB) when adjustable
and intentional noise is introduced into encrypted data in application
layer. A verification strategy in exhaustive search step of linear
attack is designed to allow Eve to mount a successful attack in the noisy
environment. Thus, a controllable wiretap channel is created over multiple
frames by taking advantage of errors in Eve's cryptanalysis, whose secrecy
capacity is found for the case of known channel states at receivers. As
a result, additional secure information can be delivered by performing
Wyner type secrecy encoding over super-frames ahead of encryption, namely,
our proposed secrecy encoding-then-encryption scheme.  These secrecy
bits could be taken as symmetric keys for upcoming frames. Numerical
results indicate that  a sufficiently large secrecy rate can be achieved
by selective noise addition.

\end{abstract}


\section{Introduction}

Traditionally, end-end secrecy delivery relies on symmetric or
asymmetric encryption residing in the upper layer of a communication
system, as well as sophisticated key management schemes \cite{Ehrsam,
Schneier:1995}. Without requiring a secure cipher, Wyner-type secrecy
encoding provides a completely different solution to link-wise secret
message delivery by random binning tailored to some presumed wiretap
channel models in physical layer \cite{Wyner,Csiszar}. In this paper,
we propose an encoding-encryption approach to end-end secrecy delivery by
encoding over a degraded wiretap channel across super-frames transmitted
in the application layer. The resulting wiretap channel is created
by injecting controllable noise into ciphertext after encryption, and
determined by both the adversary node's uncertainty about the key of
cipher and its limited resources in launching cryptanalysis. Secrete
information transmitted in such manner could be taken as keys for the
subsequent super-frame.

In the proposed framework, we are essentially exploring the techniques
developed for physical layer secrecy encoding and cryptanalysis against
symmetric block ciphers to serve our purpose of realizing end-end
secrecy enhancement  without resorting to exogenous physical channel
conditions. More specifically, Data Encryption Standard (DES) block cipher
working in Cipher Feedback Mode (CFB) is taken to encrypt messages encoded
using the Wyner type secrecy encoding scheme  and then transmitted over
multiple frames encrypted using different keys.  Random binary noise
is then deliberately added onto ciphertext, which are received by both
legitimate user and an eavesdropper without any additional distortion.
Such a hierarchical encoding-encryption framework allows us to transmit
secrete messages over the resulting degraded wiretap channels in the
application layer without making any assumption regarding  end-end
physical channel conditions.

In order to analyze secrecy enhancement achieved by utilizing our
encoding-then-encryption approach, we need to study how Eve responds to
the existing noise in her gathered data, and how that influences her
cryptanalysis performance.  In our case, Eve attempts  to mount her
linear attack with accumulated noisy ciphertexts, and thus  applies
a new verification strategy in the second phase of the linear attack
while considering her possible resource constraints. Our statistical
analysis shows that even when she uses a numerically optimized
attacking strategy to obtain the key, it is likely for her to make
mistakes in cryptanalysis. These possible failures of Eve over multiple
frames make her channel degraded than the main channel, which can be
further exploited by secrecy encoder to send additional secret bits over
a super-frame.  Therefore we could utilize generated secret bits over
the last super-frame, whose secrecy is ensured by Wyner-type secrecy
encoding scheme, to establish keys for next coming frames. The secrecy
capacity of the system is computed assuming known channel states
at Bob and Eve. Numerical results illustrate how deliberately added
noise influences secrecy rate which can be further maximized at certain
noise rate. It should be noted that the primary goal of our paper is to
demonstrate through such a case-study how secrecy encoding and symmetric
encryption could be put together to enhance end-end security, and thus
we only provide capacity computation of the resulting channel towards
the end without dealing with the implementation of a particular secrecy
encoder \cite{Thangaraj}.

In literature, very few analytical approaches have focused on the
impact of noisy ciphertexts on the attacking performance. In \cite{Yin}
different security schemes are analyzed from both reliability and
secrecy perspectives in the presence of channel noise; nonetheless,
they do not discuss what modified strategy Eve needs to take adaptively
against degradation, and nor have they considered further leveraging
adversary's failures in its cryptanalysis.  In fact, our approach shares
a common spirit with friendly jamming schemes proposed in physical layer
secrecy encoding \cite{Goel,Vilela} where deliberate noise is introduced
in physical layer to interfere both legitimate link and eavesdropped link
to  improve the secrecy rate region. Unlike these works where link-wise physical
channel features are explored to create a degraded wiretap channel, we
essentially explore the adversary's disadvantages due to its uncertainty
about the secrete key bits and resulting deteriorated success rate
in cryptanalysis in the presence of deliberate noise.  

In addition, deliberate additive noise in encryption process was used
to improve security of ciphers  in previous works \cite{Mihaljevic,
Willett, Mihal}.  The primary goals in these works were to enhance the
secrecy of a cipher by random binning and additive noise, not the one we
are interested in, namely, deploying encoding-then-encryption framework
to enhance secrecy by further encoding over a resulting degraded
wiretap channel.  Random measurement noise has also been considered
in side channel attacks (SCA) where information about cryptographic
operation is leaked through some physical measurements conducted by  an
adversary \cite{Kocher:1999:DPA}.  In \cite{Roche}, authors proposed to
use multi-linear approximation utilized in Differential  Power Analysis
(DPA)-like attacks, which is powerful due its robustness against noise,
to attack a symmetric cipher hardware by power analysis

The paper is organized as follows. In sections \ref{sec.preview},
a preliminary description of CFB mode and linear cryptanalysis
is provided. In section \ref{sec.scheme} the proposed security
scheme is described in detail,  and in \ref{sec3}, we design an
optimized verification strategy for Eve. In section \ref{sec7} the
main channel and wire-tap channel are modeled and then the secrecy
capacity of the resulting degraded wiretap channel is found in section
\ref{sec10}. Finally, we present the numerical results in section
\ref{sec15} and conclude the work in section \ref{sec16}.

\section{Review of Relevant Background}\label{sec.preview}

\subsection{Properties of CFB Mode of Operation with DES Cipher}\label{sec1}

\begin{figure}
\centering
\includegraphics[width=3.2in]{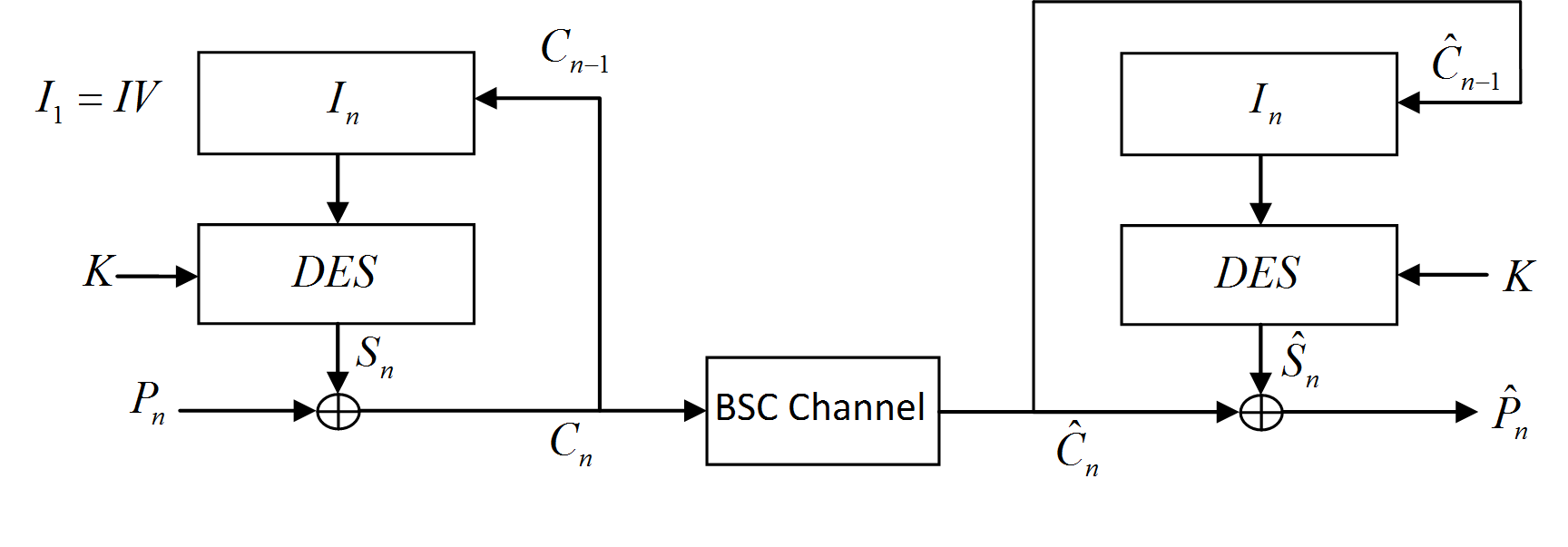}
\caption{Cipher feedback mode (CFB) with DES cipher \cite{Yin}} \label{fig1}
\end{figure}

DES is a symmetric key encryption cipher which has plaintexts and
ciphertext of size 64-bit with the key length of 56 bit. Although DES is
replaced by AES in some applications, it is still used and studied in
many networks \cite{Yijun} and \cite{Zibideh}. CFB mode is one of the
operational modes that can be used to derive a key stream from block
ciphers like DES \cite{Xiao}.  We assume that block size in CFB mode
is 64-bit. As can be seen in Fig. \ref{fig1}, at time $n$, encryption
of previous ciphertext block $C_{n-1}$ generates the key stream $S_n$
which Xors with the the current 64-bit plaintext $P_n$, to generate
64-bit ciphertext block $C_n$, i.e. $C_n=S_n \oplus P_n$ where $S_n =
E_K(C_{n-1})$.

DES encryption is very sensitive to the noise introduced into ciphertexts or key bits. In particular, when one bit of the key or the input to the cipher is altered, it can deteriorate about half of the cipher output. This property is called avalanche effect \cite{Heys1}. However, since S-boxes in DES cipher are not ideal, the resulted bit error rate by avalanche effect is not exactly $0.5$. This is also true in more advanced ciphers like AES as discussed in \cite{Nyberg}. That is why in our analysis we assume that when there is an error in cipher input or in the key, each output bit is flipped with the probability of $\alpha$.

\subsection{Linear Cryptanalysis}\label{sec20}

Linear cryptanalysis is a known plaintext attack which was first
proposed by Matsui in \cite{Matsui1} to attack DES. It is one of the
most widely used attacks on block ciphers. This cryptanalysis approach
exploits a linear equation with the probability of $p\neq\frac{1}{2}$
which involves some input and output bits of the DES cipher and some key
bits. The quantity $\varepsilon=|p-\frac{1}{2}|$, which is called bias,
measures the correlation among plaintext, ciphertext and the key bits,
and can be used as a criterion to distinguish the right key. Before
attack, Eve has to gather a large number of plaintext/ciphertext pairs,
and then for each possible key value compute its corresponding bias by
counting the number of pairs that satisfy the linear equation.

If we refer to $m$ as the number of attacked key bits in linear cryptanalysis, the number of subkey candidates would be $2^m$ that need to be sorted from rank $1$ to $2^m$ based on their corresponding probability biases. It should be noted that it is not necessarily always true that the right key ranks the highest, but it will be surely among high ranked candidates. Assume that adversary only checks top $2^{m-a}$ candidates during exhaustive search, and since each subkey candidate gets checked with all possible combinations of $56-m$ remaining unattacked bits, Eve has to run exhaustive search with at most $2^{56-m}$ encryptions for each candidate. As a result, the total number of 56 key bits examined in linear attack with bit advantage $a$ is $2^{56-a}$. In \cite{Selcuk}, A. Sel\c{c}uk showed that when the total number of gathered plaintext/ciphertext pairs $N$ are large enough, the probability of success $P_s$, defined as the probability that the right key is among $2^{56-a}$ top candidates, can be derived as
\begin{gather}
P_s=\Phi(2\sqrt{N}\varepsilon-\Phi^{-1}(1-2^{-a-1}))\label{eq},
\end{gather}
where $a$ is the bit advantage of the attack, $\varepsilon$ is the bias of the used linear approximation and $\Phi$ is the cumulative distribution function of the standardized normal distribution.

\section{The proposed scheme for security system}\label{sec.scheme}

\begin{figure}
\centering
\includegraphics[scale=0.45]{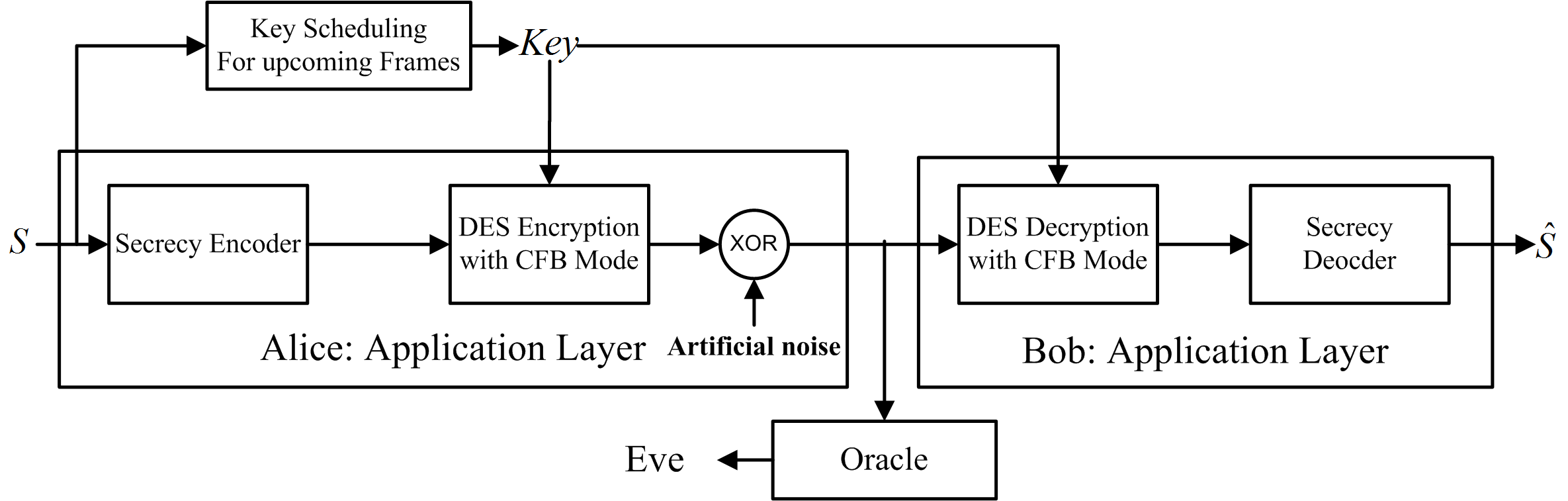}
\caption{The proposed security scheme based on the intentional noise} \label{scheme}
\end{figure}

Fig. \ref{scheme} illustrates the proposed scheme for secrecy
improvement in which after encryption of the original message $S$,
intentional noise is injected into it to generate a degraded wiretap
channel. Since we consider end to end secrecy, physical channel is assumed
to be error-free. Therefore, the ciphertexts that Bob obtains only
include errors caused by intentional noise introduced into encrypted
data in application layer with bit error rate of $\eta$. Moreover,
because Alice and Bob agree on the key used for the current data
frame, Bob can decrypt the obtained noisy ciphertexts and then apply
the wiretap channel decoding algorithm that allows him to recover the
original message $\hat{S}$ with arbitrarily small error probability. As
indicated in Fig. \ref{scheme}, there exists an oracle which is located
after encryption and noise injection, whereby Eve can query and obtain
consecutive plaintext/ciphertext pairs. However, due to the deliberate
noise, the virtual oracle provides Eve with noisy ciphertexts distorted
by a binary noise sequence with independent errors of rate $\eta$.
The main advantage that Bob has over Eve is that Bob and Alice share the
same encryption and decryption key which is unknown to Eve. Therefore,
Eve has to adopt an attack strategy that can exploit the gathered noisy
data in order to guess the secret key.

We assume that legitimate users initialize with a shared set of keys in
a highly secure manner at the beginning. As a result, Alice can divide
the whole data into equal size data frames, each including $M$ number of
data blocks of size $64$-bit which is the block size used in CFB mode. In
this way, the same key will be used for $M$ $64$-bit blocks in each
frame for encryption and decryption at the receiver end. In this paper,
we show that due to Eve's resource constraints, it is likely for her to
make mistakes in assessing a frame key. As a result, Eve's channel is
a degraded version of the main channel. We can leverage this advantage
by applying Wyner secrecy encoding over super-frames to average over
all possible failures by Eve. In Wyner-type encoder redundancy is added
to correct errors that occur across the main channel, and randomness is
added for keeping Eve ignorant across the wiretap channel \cite{Wyner},
\cite{Thangaraj}. Note that this scheme can be generalized for other block
ciphers like AES when they are used in operational modes like CFB or CBC
(Cipher Block chaining).

Another issue is key scheduling problem to provide highly confidential and
distinctive keys for each frame while Bob is fully aware of them. Here,
we can use traditional way of key management which is sophisticated
and costly. For instance, master/session key scheduling approach which
is proposed for DES cipher in  \cite{Ehrsam, Schneier:1995}. In this
technique, there exists a master key out of which frame keys as session
keys can be originated. In our scheme, we propose a simpler approach
which requires less expenditure. In this technique the secrecy required
for frame keys is originated from secret bits delivered by Wyner secrecy
encoder over the intentionally created wiretap channel. As a result,
since encoder is performed over each super-frame,  Alice can use input
to the encoder to derive frame keys in next super-frame, for instance
by applying a universal class of Hash functions \cite{Bennett}, where
the utilized function for each frame is publicly known. Bob is able to
decode encrypted data and obtain the encoded message, and thus he will
be able to derive keys for next frames. Note that the requirement for
this approach is that there has to exist some root keys to initiate the
keys for the first super-frame.

\section{Eve's attack strategy and its analysis in noisy environment}\label{sec3}

This section studies the effect of the channel degradation on the performance of the linear cryptanalysis in terms of Eve's success rate. Since linear cryptanalysis is a known plaintext attack, Eve has to rely on the received plaintext/ciphertexts pairs. Due to the existing errors in these ciphertexts, when Eve examines a key, she is unable to distinguish between errors caused by the received noisy ciphertext and the ones induced by using the wrong key. Thus, she needs to design a new verification approach whereby she can find the right key. It should be noted that this attack strategy with verification process has to be designed in a way that attack success rate gets maximized from Eve's perspective. 

\subsection{Designed Verification Strategy for Attack}\label{sec4}

Consider ciphertexts go through a binary symmetric channel whose cross-over probability is $\eta$. As seen in Fig. \ref{fig1}, after $C_n$ passes through channel, and Xors with channel noise, the received noisy $64$-bit ciphertext $\hat{C}_n$ will have error with the probability of $1-(1-\eta)^{64}$. Therefore, Eve can not rely only on two successive ciphertexts to check the correctness of a key, because they might have errors that can lead her to make mistakes. Indeed, Eve has to try a number of successive pairs, using CFB mode in order to increase her success rate.

\begin{figure}
\centering
\includegraphics[scale=0.4]{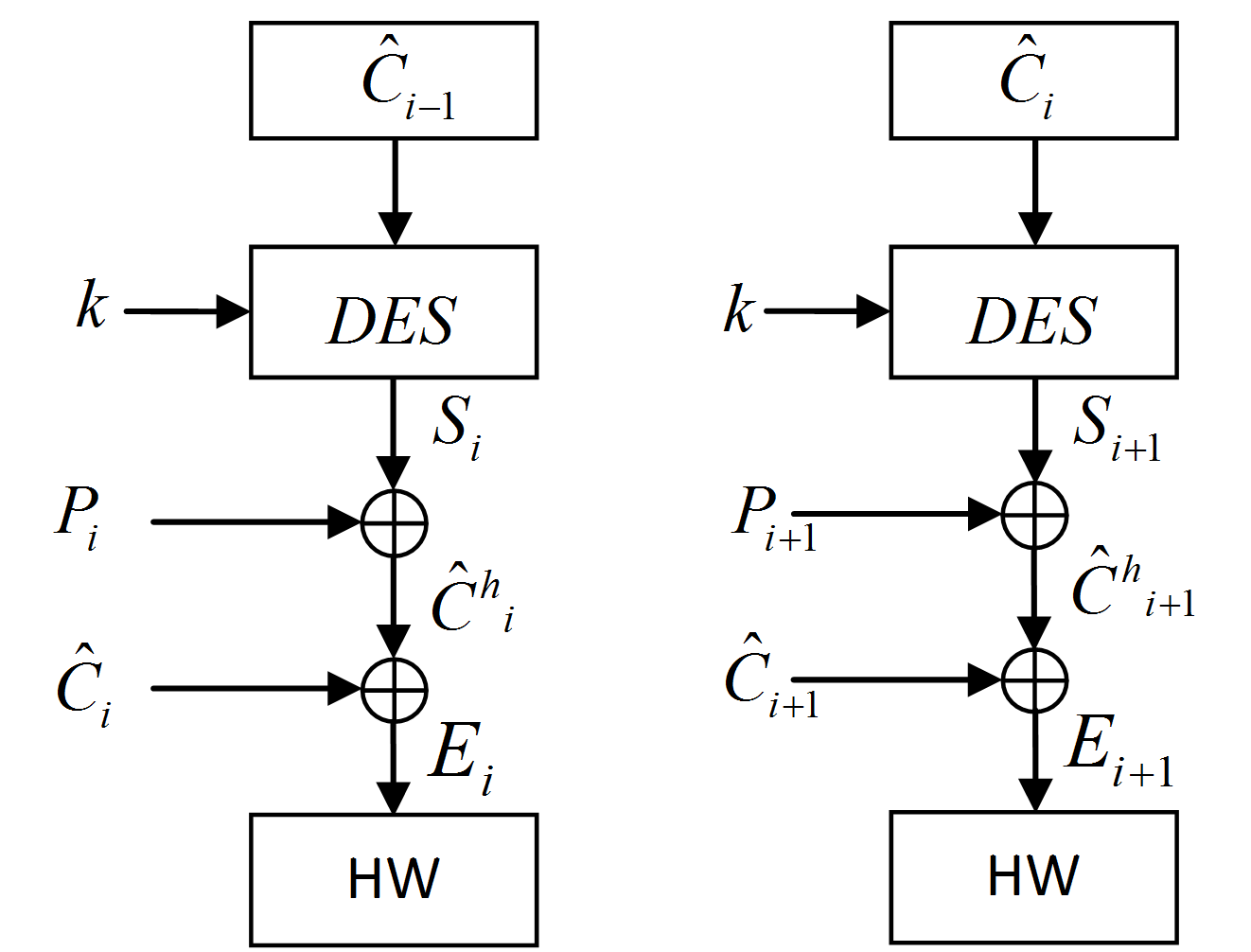}
\caption{key verification process for Eve with two consecutive CFB stages} \label{fig2}
\end{figure}

In Fig. \ref{fig2}, two consecutive stages of CFB that are used to check the key are shown, where $P_i$ and $\hat{C}_i$ are respectively the plaintext and ciphertext for the $i^{th}$ stage, $S_i$ is the encrypted result of $\hat{C}_{i-1}$ that after Xor with $P_i$ generates $\hat{C}^h_i$. Provided that the used key is correct, $\hat{C}^h_i$ must be the same as $\hat{C}_i$. However, due to the possible errors in $\hat{C}_i$ or $\hat{C}_{i-1}$ there might be some differences between $\hat{C}_i$ and $\hat{C}^h_i$ even though the used key is right. Therefore, Hamming Weight (HW) of Xor of $\hat{C}^h_i$ and the ciphertext $\hat{C}_i$ denoted by $E_i$ must be compared with a threshold denoted as $\tau$. Then, a key trial for the $i^{th}$ stage can be considered successful if this HW is less than $\tau$.

Note that at stage $i$ when there is an error either in the input to the cipher, i.e. $\hat{C}_{i-1}$ or in the key, there will be burst of errors in $S_i$, which makes $\hat{C}^{h}_{i}$ totally different and in special case of $\alpha=0.5$ independent from $\hat{C}_i$. Therefore, by choosing a small value for threshold $\tau$ and comparing HW of $\hat{C}^h_i \oplus \hat{C}_i$, Eve can know that either input to the cipher or the key is noisy. In Table \ref{attack}, the key verification strategy for Eve is given that she needs to follow in the brute-force attack phase of linear cryptanalysis to test the correctness of the examined key $k_i$. In this strategy, Eve examines each key candidate $N_c$ times with $N_c$ consecutive pairs. When at least one of trials is successful, Eve decides that the key is correct. That is because for a correct key, $N_c$ is chosen such that she can make sure that with a high probability at least in one trial out of $N_c$ tests, input to the cipher has no error that results in a success.

\begin{table}[h!t!p!]
\caption{Verification strategy}
\begin{tabular}{l}
\hline
1- Pick $N_c$ number of consecutive pairs.\\
2- Try $N_c$ chosen pairs over $N_c$ chained CFB stages using the key $k_i$.\\
3- A trial is successful if $HW(E_i=\hat{C_i}\oplus \hat{C}^h_i)\leq \tau$.\\
4- If there exists at least one successful event out of $N_c$ trials,\\
 \quad $k_i$ is the correct key, otherwise it is wrong.\\
\hline
\end{tabular}
  \label{attack}
\end{table}

Now the question is how we can choose the optimum value for $\tau$. When the tested key is right, at stage $i$, for error-free $\hat{C}_{i-1}$, $\hat{S}_i$ will be error free and all the errors in $\hat{C}^h_i$ will be caused by the possible errors in $\hat{C}_i$. However, we can choose $\tau$ such that with a high probability, the number of errors in $\hat{C}_i$ does not exceed this threshold. Hence, the minimum possible value for $\tau$ has to be adjusted such that at stage $i$, the probability that the number of bit errors in $\hat{C}_i$ exceeds $\tau$ becomes negligible. This probability is denoted by $P_{fault}$
\begin{gather}\label{eq9}
P_{\emph{fault}}=1-\sum_{i=0}^{\tau} \binom{64}{i}(\eta)^i(1-\eta)^{64-i}.
\end{gather}

In the next step, we need to find the optimum value for $N_c$. Suppose that Eve tries a key to see if it is the right one, and let $K^h_0$ be the hypothesis when the key is wrong and $K^h_1$ when it is right. Then, we introduce random variable $A_i$ such that $A_i=1$ defines successful trial at the $i^{th}$ stage that happens when Hamming weight of $E_i$ is less or equal to $\tau$, and $A_i=0$ otherwise. By proper selection of $\tau$, We can make sure that whenever there is no error in the input to the cipher, Eve can recognize the right key. Hence, the probability of having a successful event at the $i^{th}$ stage given the right key will be
\begin{gather}\label{eq.p1-2}
P_1=Pr[A_i=1|K^h_1]=P[\hat{C}_{i-1} \textrm{is error-free}]=(1-\eta)^{64}.
\end{gather}

All $N_c$ tests will fail if in all of these trials, inputs to the ciphers have error. If it happens when the key is right, Eve will miss it, which has the probability of
\begin{gather}\label{eq11}
P_m = (1-P_1)^{N_c},
\end{gather}
We call $P_m$ key missing probability. Thus, we need to find minimum $N_c$ such that keeps $P_m$ below a threshold like $T_m$.

Now we need to compute the probability that Eve mistakenly admits a wrong key while examining a single candidate. When the used key is wrong due to the avalanche effect, $\hat{C}^h_i$ will have bit error rate of $\alpha$, that after Xor with $\hat{C_i}$ with bit error probability of $\eta$, results in output bit error rate of $\gamma$ as
 \begin{gather}\label{eq.gama}
 \gamma=\alpha (1-\eta)+\eta(1-\alpha).
 \end{gather}
Since to admit a wrong key at the $i^{th}$ stage as the right one, HW of $E_i$ must be less than $\tau$, the probability of a successful trial at this stage for a wrong key is
\begin{gather}\label{eq8}
P_2=Pr[A_i=1|K^h_0]=\sum_{i=0}^{\tau}\binom{64}{i}\gamma^i(1-\gamma)^{64-i}.
\end{gather}

On the other hand, Eve accepts a wrong key when there happens at least one successful trial for it. Thus, the false key probability for a single candidate is
\begin{gather}\label{eq12}
P_F = 1-(1-P_2)^{N_c},
\end{gather}
where $P_2$ is computed by Eq. (\ref{eq8}). It may seem that $P_F$ is very negligible for the case $\alpha=0.5$ in which $\gamma=0.5$. However, this probability can be aggregated over a large number of examined wrong candidates in linear attack, and can lead to an overall false key probability that can not be neglected, as will be seen in simulation results.

\subsection{Analysis of the Designed Attack Strategy for Eve}\label{sec5}

In \cite{Yin} Yin et. al. showed that in noisy environment with bit error rate of $\eta$, for linear attack on DES cipher, the probability bias of the new linear equation denoted by $\hat{\varepsilon}$, as well as the success probability of attacker $P_s$ can be computed based on the linear probability bias of the original linear equation $\varepsilon$ and the number of obtained pairs by Eve $N$ as
\begin{align}\label{eq17}
&P_s=\Phi(2\sqrt{N}\hat{\varepsilon}-\Phi^{-1}(1-2^{-a-1})),\nonumber\\
 &\textrm{where}\quad\hat{\varepsilon}=2^{u+v}(1-\eta-0.5)^{u+v}\varepsilon.
\end{align}
If adversary uses the improved linear analysis technique, she needs to use Matsui's linear equation for DES that requires $u$ bits of plaintext and $v$ bits of corresponding ciphertext where $u+v=26$ to guess $m=26$ key bits \cite{DESEXP}. As discussed in subsection \ref{sec20}, in linear attack with bit advantage of $a$, the total number of examined keys is $2^{56-a}$.
If the ciphertexts that Eve obtains are error-free, her success probability will be $P_s$ which is the probability that the correct key is among top $2^{56-a}$ examined candidates. However, when her obtained ciphertexts are erroneous, it is still likely for her to obtain the frame key. Also, it is possible that she gets no frame key either right or wrong for decryption, which imposes her to erase the whole frame. These events have probabilities that are called total success probability and frame erasure probability, respectively, that can be computed based on the following theorem which is proven in appendix \ref{appendix.a}.

\vspace*{4pt}
\begin{thm}\label{total_success}
\textit{Consider a linear attack with bit advantage of $a$. Assume Eve's obtained ciphertexts contain bit errors with the rate of $\eta$, and that she uses the designed strategy in brute-force step of the linear attack. When Eve examines the right key, she misses it with the probability of $P_m$ given in Eq. (\ref{eq11}), and when the key is wrong, she may accept it wrongly with the probability of $P_F$ given in Eq. (\ref{eq12}). Let $P_s$, given in Eq. (\ref{eq17}), be the success probability when the ciphertexts are error-free. Then, Eve's total success probability can be computed by
\begin{align}\label{eq.pc}
P_c=\frac{P_s(1-P_m)}{P_F 2^{56-a}}[1-(1-P_F)^{2^{56-a}}]\approx P_s(1-P_m),
\end{align}
On the other hand, frame erasure probability will be
\begin{align}\label{eq.pe}
P_e &= (1-P_s)(1-P_F)^{2^{56-a}}+P_s P_m(1-P_F)^{(2^{56-a}-1)}\nonumber\\
&\approx [1-2^{56-a}P_F][1-(1-P_m)P_s].
\end{align}}
\textit{In addition, the probability that Eve accepts a wrong key in linear attack which we call wrong key probability denoted by $P_w$ can be derived as $P_w=1-P_c-P_e$.}
\end{thm}
\vspace*{4pt}
Conclusively, we showed that there is possibility that Eve is not able to obtain any key, or to falsely accept a wrong key.

\subsection{Parameter Optimization of Adversary's Attack Strategy}\label{sec6}
Eave's objective is to mount a successful attack, and in order to achieve this goal, she maximizes the success probability of the utilized linear attack $P_c$, given in (\ref{eq.pc}), knowing that her computational ability is restricted, and there is a constraint on the number of plaintext/ciphertext pairs that she can accumulate. Namely, she can not perform more than $\theta$ DES encryptions. In the linear cryptanalysis designed for noisy environment, the number of all examined keys is $2^{56-a}$ and each one has to be checked for $N_c$ times. Hence, in the worst scenario Eve has to run $N_c 2^{56-a}$ DES encryptions, which due to Eve's computational restrictions, should not exceed $\theta$. Moreover, we assume that before mounting attack on a frame of data, Eve has already gathered as many number of pairs as data storage capability and time limit allow her denoted by $N_{max}$. As a result, Eve needs to design attack parameters including $N_c$, $\tau$ and $a$, to maximize the overall success probability subject to the following constraint
\begin{gather}\label{eq24}
\displaystyle\max_{N_c,\tau,a} P_c\qquad \textrm{subject to}~~\theta \geq N_c . 2^{56-a}, \quad N \leq N_{max}.
\end{gather}

\begin{table}[h!t!p!]
\caption{Parameter optimization algorithm for attack strategy:}
\begin{tabular}{l}
\hline
1- Initialization: put $\tau=1$, $N_c=1$. \\
\quad Determine $T_m$ and $T_f$ as thresholds for $P_m$ and $P_{\emph{fault}}$\\
\quad also $N_{cmax}$ as the maximum value for $N_c$.\\
2- $\tau\leftarrow \tau+1$  until $P_{\emph{fault}} > T_f$ and $\tau<64$\\
\quad if $P_{\emph{fault}} \leq T_f$ or $\tau=64$ go to the next step\\
3- $N_c\leftarrow N_c+1$ until $P_m > T_m$ and $N_c<N_{cmax}$\\
\quad if $P_m \leq T_m$ or $N_c=N_{cmax}$ go to the next step\\
4- Compute $a_0=\lceil 56-\log_2(\frac{\theta}{N_c}) \rceil$\\
5- Compute $P_c$ for $a_0\leq a\leq 56$\\
\quad choose $a$ for which $P_c$ has its largest value.\\
6- Output $\tau$, $N_c$ and $a$ as attack parameters.\\
\hline
\end{tabular}
  \label{tab1}
\end{table}

From Eq. (\ref{eq.pc}) it can be concluded that $P_c$ falls as $P_m$ increases. Since according to Eq. (\ref{eq11}), $P_m$ mainly depends on $N_c$, we can define threshold $T_m$ for it and find the minimum number of trials $N_c$ for which $P_m$ remains below $T_m$. According to equations (\ref{eq8}) and (\ref{eq12}), to decrease $P_F$ we need to reduce $\tau$ as much as possible. If we define a threshold $T_{f}$ for $P_{fault}$, the minimum possible value for $\tau$ according to our discussion in \ref{sec4} is the smallest $\tau$ for which $P_{fault}$ remains below $T_f$. Furthermore, Eve has to choose an optimized value for $a$ to have $P_c$ maximized. The algorithm in Table \ref{tab1}, is designed to optimize the linear attack parameters to let Eve achieve the maximum success rate $P_c$, for a given $\eta$ subject to her restrictions. In this algorithm, $P_{\emph{fault}}$ and $P_m$ can be computed using equations (\ref{eq9}), (\ref{eq11}), respectively.

\section{Main and Wire-tap Channel modelling}\label{sec7}

In this section, we model main and wiretap channels in block level (with $64$-bit input and $64$-bit output), using a stationary finite state Markov chain (MC). Since Eve might achieve the right frame key, get a wrong one or even get nothing and drop the whole frame, we also need to model her channel in frame level as a three state memoryless channel.

\subsection{Main Channel Modelling Using MC}\label{sec8}

As it was described, the encrypted data goes through a BSC channel with cross over probability of $\eta$, created by intentionally introduced noise in application layer. We next model the CFB cipher, channel with deliberate noise and decipher altogether as a single channel, in order to analyze the effect of intentional noise at the output of decipher. Note that we assume there is no degradation in actual physical channel.

\begin{figure}
\centering
\includegraphics[width=2.6in]{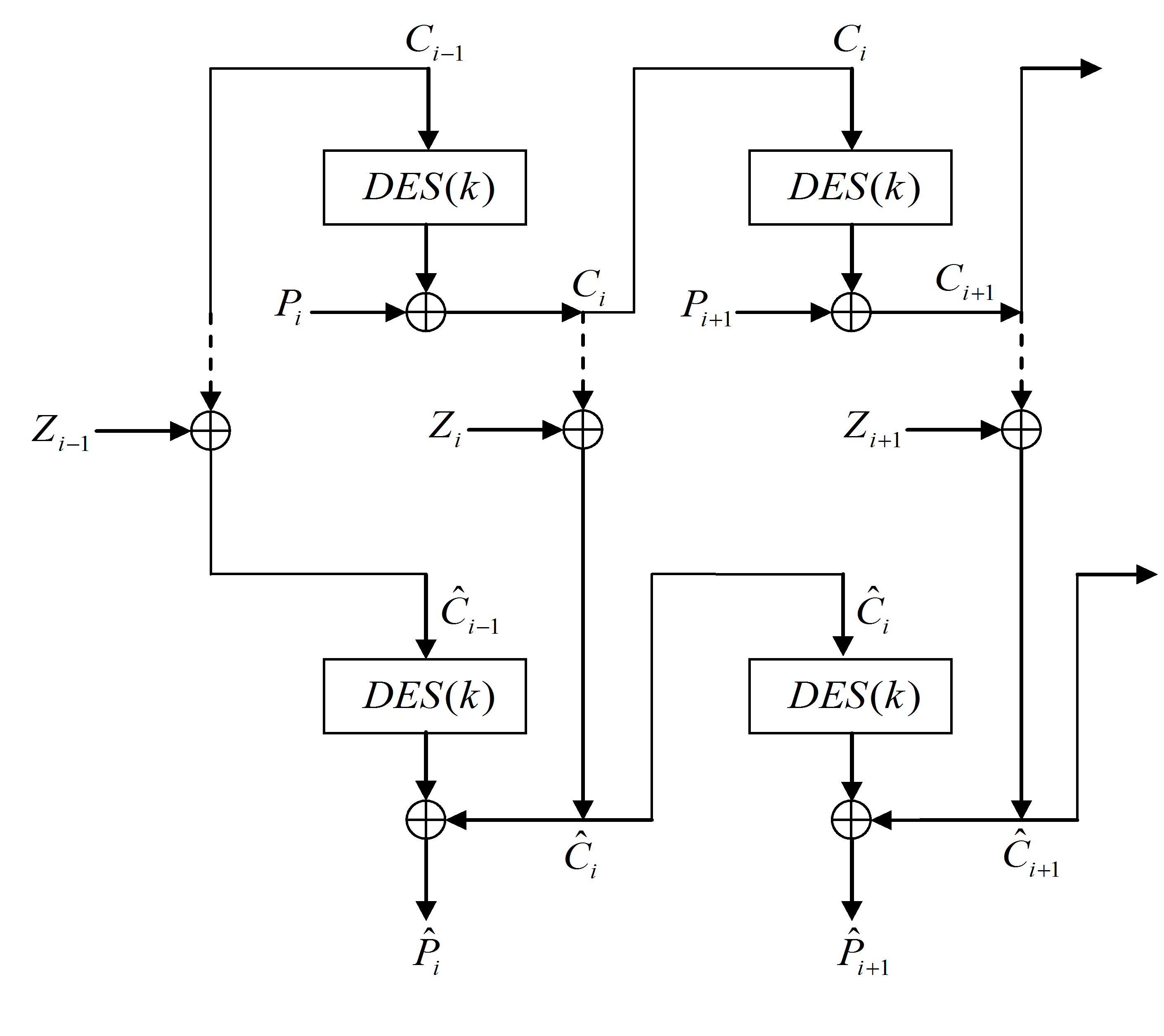}
\caption{CFB enciphering and deciphering with channel error}\label{fig.encdec}
\end{figure}

Fig. \ref{fig.encdec} illustrates the encryption and decryption structure of CFB mode with DES cipher in the presence of introduced noise to ciphertexts. As shown in this figure, $\{C_i\}$ and $\{\hat{C}_i\}$ are the sequences of transmitted 64-bit ciphertext and received noisy ciphertext blocks, respectively, and $\{\hat{P_i}\}$ is the sequence of decrypted blocks at time $i$ for $i=1,2,\ldots$. In addition, $\{Z_i\}$ is the sequence of 64-bit blocks of intentional bit errors in channel $Z_i^j$ that are independent and identically distributed with Bernoulli distribution as $Pr[Z_i ^j=1]=\eta$ for $j=1,\ldots,64$, such that $\hat{C}_i=C_i \oplus Z_i$. As Fig. \ref{fig.encdec} indicates when $\hat{C}_i$ is noisy, it introduces errors with the rate of $\eta$ to the decryption output at time i, i.e. $\hat{P_i}$. Moreover, since $\hat{C}_{i-1}$ gets encrypted with DES at time $i$, due to the avalanche effect, it induces bit error rate of $\alpha$ in $\hat{P}_{i}$. As a result, to characterize the channel error state in decryption output at time $i$, it is required to consider errors in both currently received ciphertext $\hat{C}_i$ and the previous one $\hat{C}_{i-1}$. Hence, we need to define four states.

Note that in a particular case when we consider $\alpha=0.5$, we still need to define four states. In this case, when $\hat{C}_{i-1}$ has error, due to the fact that half of the ciphertext will be in error, errors in $\hat{P}_{i}$ will be independent from $\hat{C}_i$ and consequently from the error state at time $i+1$. However, when it has no error, errors in $\hat{C}_i$ will affect both decryption outputs at times $i$ and $i+1$, and therefore the current state will depend on the previous one. As a result, we have to take all four states into account, each with a different transition probability from the input plaintext block $P_i$ denoted as $64$-bit vector $X$ to the output stored plaintext $\hat{P_i}$ denoted by $64$-bit vector $Y$, and let $E=X\oplus Y$ denotes the transition error vector.

The channel states are defined as: state $S_0$, in which there is no error from vector $X$ to the vector $Y$ and happens when there is no error in $\hat{C}_i$ and $\hat{C}_{i-1}$. State $S_1$, which happens when there is at least one bit error in $\hat{C}_i$, but no error in DES cipher input, $\hat{C}_{i-1}$. State $S_2$, which shows the situation in which there is at least one bit error in $\hat{C}_{i-1}$ without any error in $\hat{C}_i$. In this channel state, due to the avalanche effect, each bit at the output of DES cipher, flips independently with the probability of $\alpha$ causing bit error probability of $\alpha$ in $Y$. State $S_3$, in which both $\hat{C}_i$ and $\hat{C}_{i-1}$ have at least one bit error.

For state $S_0$ we have $Pr[e_j=1|S_0]=0$ and for $S_2$, $Pr[e_j=1|S_2]=\alpha$, where $e_j$ denotes the $j^{th}$ bit of $E$ for $j=1,\ldots,64$. On the other hand, we should note that in states $S_1$ and $S_3$, output bits can not be treated independently because $S_1$ and $S_3$ are based on a given condition on the whole $64$-bit ciphertext $\hat{C}_i$. Let $q$ denote the probability that there exists at least one bit error in $Z_i$ as
\begin{align}\label{eq.ers1c}
q=1-(1-\eta)^{64}.
\end{align}
The next lemma gives the input-output transition probability for states $S_1$ and $S_3$, which is proven in Appendix \ref{ap.B}.

\begin{lem}
\label{Vector transition probability}
\textit{Let $X$ be the input plaintext vector to the CFB encryption mode and $Y$ be the corresponding output of the decryption. If the generated ciphertexts go through a channel with cross over probability of $\eta$, we denote the Hamming weight of the resulted error vector $E$ with $W(E)$. Then, for state $S_1$ the input-output vector transition probability will be
\begin{align}\label{eq.ers1}
Pr(Y|X,S_1)=\left\{\begin{array}{ll} \frac{\eta^{W(E)}(1-\eta)^{64-W(E)}}{q} & W(E)\neq 0 \\
0 & W(E)= 0
\end{array} \right.
\end{align}
where $\alpha$ is the avalanche bit error rate, and $\gamma$ is given in Eq. (\ref{eq.gama}). The transition probability in state $S_3$ for all $W(E)$ is
\begin{align}\label{eq.ers3}
&Pr(Y|X,S_3)=\\
&\frac{\gamma^{W(E)}(1-\gamma)^{64-W(E)}-\alpha^{W(E)}(1-\alpha)^{64-W(E)}(1-q)}{q}.\nonumber
\end{align}}
\end{lem}

Next, we need to find state transition probabilities. For instance, when the state at time $i-1$ was $S_2$, apparently $\hat{C}_{i-1}$ has been error free, so the only condition required to have state $S_0$ happen at time $i$ is to receive error free $\hat{C}_i$ which has the probability of $1-q$ that is the transition probability from state $S_2$ to $S_0$. Similarly, we can compute other state transition probabilities.
\begin{figure}
\centering
\includegraphics[width=1.8in]{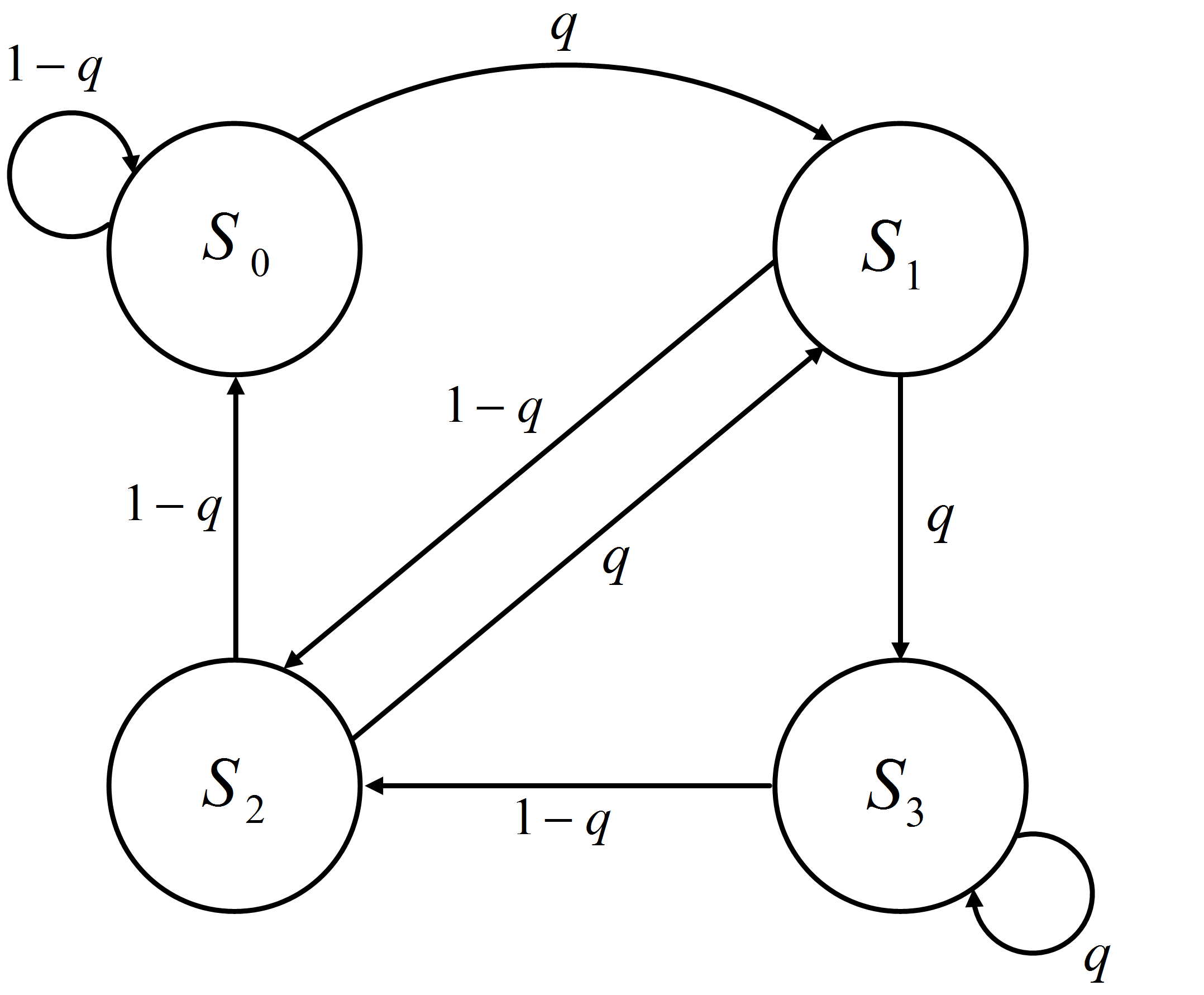}
\caption{Alice-Bob channel model as a four state MC} \label{fig4}
\end{figure}
Notably, since probability of occurrence of the current state only depends on the previous state, Bob's channel can be modeled as a four state MC that is depicted in Fig. \ref{fig4} with the following state transition probability matrix:
\begin{equation*}\label{eq33}
T=\begin{bmatrix}
1-q & q & 0 & 0 \\
0 & 0 & 1-q & q \\
1-q & q & 0 & 0 \\
0 & 0 & 1-q & q
\end{bmatrix},
\end{equation*}
whose elements demonstrate the transition probabilities between different states. Note that in each state, input plaintexts undergo different channel conditions and error probabilities. In fact, the main channel can only be modeled as a BSC channel in states $S_0$ and $S_2$ with cross over probabilities of $0$ and $\alpha$ respectively, whereas in other two states it can be modeled based on input-output transition probabilities in (\ref{eq.ers1}) and (\ref{eq.ers3}).

In particular, since in MC model for Alice-Bob channel, all four states can be reached from one another, it is an irreducible MC with positive recurrent states \cite{Ross}. Then, with a supposedly large frame size, MC can reach its stable condition. Since all states are positive recurrent, the set of equations $\textbf{P}^t \textbf{T}=\textbf{P}^t$, and $\textbf{P}^t.\textbf{1}=\textbf{1}$ have a unique solution as $\textbf{P}^t=[p_0,\ldots,p_3]$ where $p_k$ denotes the steady state probability of state $S_k$ for $k\in\{0,1,2,3\}$ \cite{Ross}. Where $\textbf{1}$ is a $4\times 1$ vector with all elements to be one, and $\textbf{P}$ is steady state probability vector (SSPV). By solving this equation set, we get
\begin{equation}
\textbf{P}^t=\begin{bmatrix}
(1-q)^2 & q(1-q) & q(1-q) & q^2
\end{bmatrix}.\label{eq.p}
\end{equation}

\subsection{Wire-tap Channel Modelling}\label{sec9}

\begin{figure}
\centering
\includegraphics[width=2.8in]{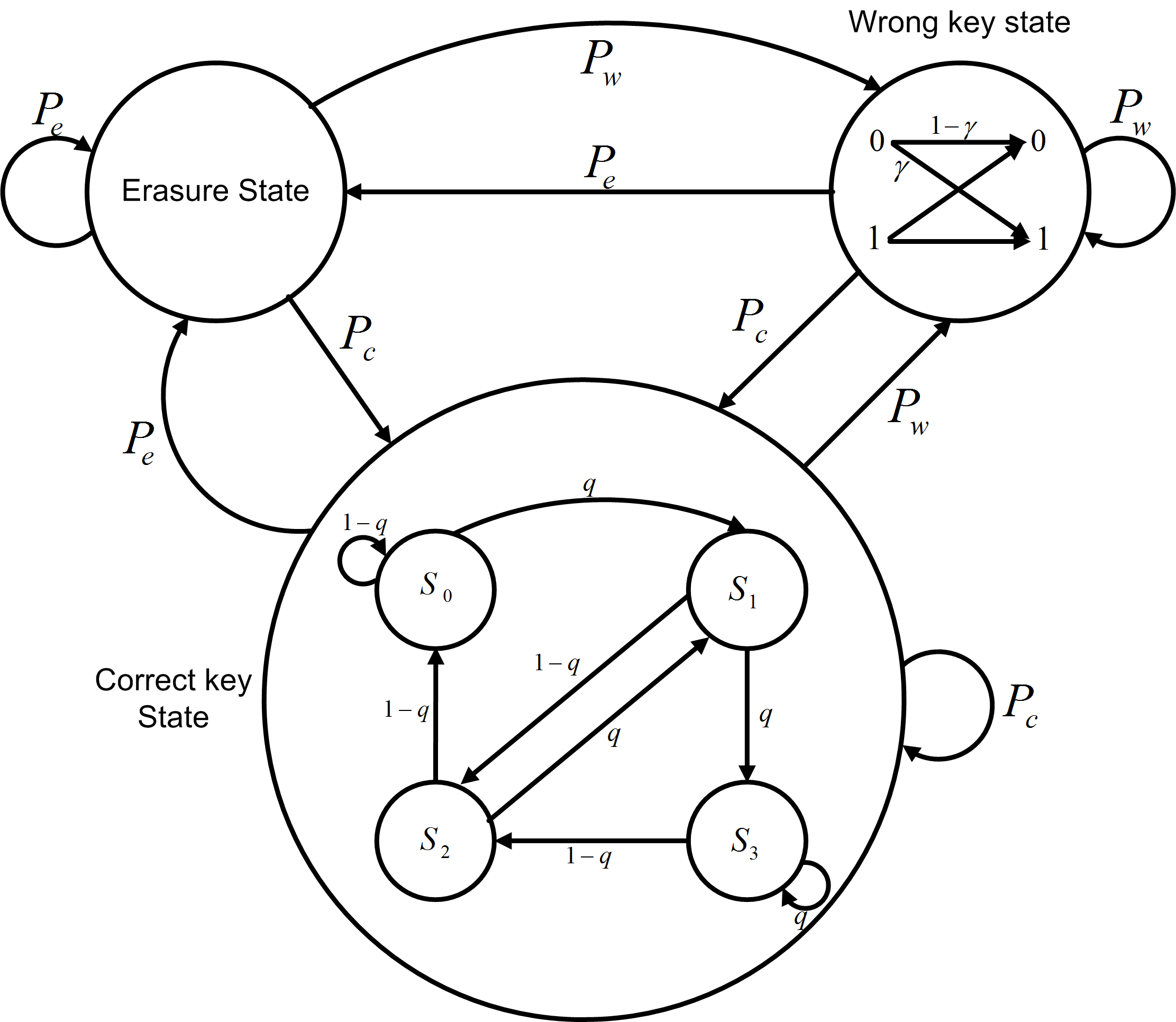}
\caption{Eve's hierarchical channel model} \label{fig5}
\end{figure}

In section \ref{sec5} we showed that adversary can obtain the right key of a frame with the probability of $P_c$ by using optimized verification strategy in linear attack. To consider the worst possible case, we assume that before starting the attack, Eve has gathered the required number of pairs such that for each frame, she has already mounted her attack. When she has been able to achieve the correct key, there will not be any difference between the main channel and her channel, so her decrypted data in that frame undergoes the same channel condition as Bob's. As shown in Fig. \ref{fig5}, we refer to this channel state for Eve as the correct key state in frame level which occurs with the probability of $P_c$ and can be modeled as a MC with four channel states in  block level.

Nevertheless, with the probability of $P_e$, Eve will not be able to get any key for the attacked frame and has to drop the whole frame. We refer to this state as erasure state. Moreover, Eve gets a wrong key with the probability of $P_w$, such that after using a wrong key due to the avalanche effect in DES cipher, each bit in DES output will be independently flipped with the probability of $\alpha$. This induced error Xors with intentional i.i.d. channel noise that has bit error probability of $\eta$. Consequently, in wrong key state, Eve's channel can be modeled as a BSC with cross over probability of $\gamma$ given in (\ref{eq.gama}). Conclusively, wiretap channel is a degraded version of the main channel that only in the correct key state can it be as good as Bob's channel. In fact, Eve's channel behaves like a pseudo two-dimensional Markov Chain (P2DMC) \cite{Yu} with three memoryless states in frame dimension, each acting like another MC in block dimension as shown in Fig. \ref{fig5}.

\section{Secrecy capacity computation}\label{sec10}

The next step is to quantify the secrecy capacity of the analyzed security system. The capacity of finite state Markov chains was calculated in \cite{Goldsmith} and \cite{Holliday}. In \cite{Wang}, \cite{Mushkin} and \cite{Lapidoth} the capacity of the finite state Markov chains with binary symmetric channels associated in each state, was studied. In \cite{yogesh} secrecy capacity of a wiretap channel modeled as a finite state MC is computed. We assume that the channel states are perfectly known to Bob and Eve, so what we compute is mutual information between the input $X$ and output $Y$ given the channel state, i.e. $I(X;Y|S_l)$. Since all four states of the main channel are in block level, in order to make Bob aware of the channel states in each block, Alice can use an error detection procedure and embed it in each block. For Eve, we assume that she is aware of this error detection procedure which allows her to beware of channel states in block level. In frame level, it is assumed that she knows the correctness state of each used frame key towards the end of each frame. Specially, this can be considered as the best scenario for Eve, providing us a lower bound for secrecy rate.

The main purpose of secrecy capacity computation is to design a secrecy encoder which is applied ahead of the encryption in application layer over multiple frames. Namely, when the message is transmitted at a rate below the secrecy rate to Bob using a Wyner-type encoding technique\cite{Ozarow}, \cite{Thangaraj}, we can have an arbitrarily small error probability for Bob as well as the maximum entropy for Eve. In the asymptotic sense, by secrecy encoding, users utilize Eve's failures which cause her channel to be a degraded channel compared to Bob's.

\subsection{Capacity of the Main Channel}\label{sec12}

When channel state information is available, the capacity is the average of capacities that each one of these MC states contribute to the overall channel capacity \cite{Goldsmith}, \cite{Wang}:
\begin{gather}\label{eq32}
C=\sum_{k=0}^{K-1}p_k C(S_k),
\end{gather}
where $C(S_k)$ is the channel capacity in state $S_k$ in bit per channel use. It can be computed as the maximum information rate between input and output vectors, $X$ and $Y$, respectively, assuming that the current state $S_k$ is known to Bob:
\begin{gather}\label{eq.capmut}
C(S_k)=\max_{P_X}I(X;Y|S_k)/64.
\end{gather}
Note that our modeled four state Markov channel is uniformly symmetric because in any state, channel is output symmetric \cite{Goldsmith}. For instance, in states $S_0$ and $S_2$, the channel behaves as a BSC channel. In states $S_1$ and $S_3$, if we define the transition probability matrix as $P_{ij} = Pr(Y=j|X=i,S_l)$ for $i \in \mathcal{Y},i \in \mathcal{X}, l=1,3$, its rows and columns are permutations of each other because according to equations (\ref{eq.ers1}) and (\ref{eq.ers3}), its elements only depend on the HW difference of input-output vectors. As a result, also in states $S_1$ and $S_3$, the channel is output symmetric. In \cite{Goldsmith} it is shown that for uniformly symmetric channel in which noise is independent of inputs, like our modeled Markov channel, capacity can be achieved with distribution which is uniform and iid. Accordingly, in this finite state Markov channel by uniformly distributed inputs, the mutual information will be essentially maximized.

In state $S_0$, channel is an error-free BSC with capacity of $1$, i.e. $C(S_1)=1$, and in state $S_2$, it acts like a BSC with cross over probability of $\alpha$ and the capacity of $C(S_2)=1-h(\alpha)$, where $h$ is binary entropy function. However, for $S_1$ and $S_3$ in which decryption bit errors are not independent, we need to compute the mutual information between input and output vectors, namely $I(X;Y|S_l)$ for $l=1,3$, that is
\begin{gather}\label{eq.mut}
I(X;Y|S_l)=H(Y|S_l)-H(Y|X,S_l).
\end{gather}
We assume that channel state is perfectly known to Bob. In the following theorem which is proven (in Appendix \ref{ap.c}) using Lemma \ref{Vector transition probability}, we compute $H(Y|X,S_l)$ for $l=1,3$.
\begin{lem}
\label{Entropy}
\textit{Consider our four state MC model for the main channel with input vector $X$ and output vector $Y$. with equally likely input plaintexts, we can compute $H(Y|X,S_1)$ as}
\begin{align}\label{eq-ent-s1}
H(Y|X,S_1)=&\frac{-1}{q}\sum_{k=1}^{64}\binom{64}{k}\eta^k(1-\eta)^{64-k}\nonumber\\
&.\log \left[ \frac{\eta^k(1-\eta)^{64-k}}{q} \right].
\end{align}
\textit{and $H(Y|X,S_3)$ will be}
\begin{align}\label{eq-ent-s3}
&H(Y|X,S_3)=\nonumber\\
&\frac{-1}{q}\sum_{k=0}^{64}\binom{64}{k}.\left[\gamma^k(1-\gamma)^{64-k}-\alpha^k(1-\alpha)^{64-k}(1-q)\right]\nonumber\\
&.\log \left[\frac{\gamma^k(1-\gamma)^{64-k}-\alpha^k(1-\alpha)^{64-k}(1-q)}{q}\right].
\end{align}
\end{lem}

On the other hand, for both states $S_1$ and $S_3$, every output vector $Y_j$ can be generated by introducing all possible error vectors over their corresponding input vectors. Hence, since all $64$-bit input plaintexts are uniformly distributed, the output will also be equally likely and uniformly distributed. Hence, for $l=1,3$ the output entropy is $H(Y|S_l)=64$. Thus, by using Eq. (\ref{eq.mut}) we can compute the mutual information for states $S_1$ and $S_3$ as
\begin{align}\label{eq.I}
I(X;Y|S_l)=64-H(Y|X,S_l),\quad \textrm{for}\quad l=1,3,
\end{align}
where $H(Y|S_1,X)$ is given in Eq. (\ref{eq-ent-s1}), and $H(Y|S_3,X)$ in Eq. (\ref{eq-ent-s3}). According to Eq. (\ref{eq.capmut}) the channel capacity in states $S_l$ for $l=1,3$ will be
\begin{align}\label{eq.cap-13}
C(S_l)=\frac{I(X;Y|S_l)}{64} \quad \textrm{(bits per channel use)}.
\end{align}
where $I(X;Y|S_1)$ and $I(X;Y|S_3)$ are given in Eq. (\ref{eq.I}). We can analyze Alice-Bob channel as a finite state MC with steady state probabilities given in Eq. (\ref{eq.p}). Hence, according to Eq. (\ref{eq32}) Bob's channel capacity $C_B$ as the average of the state capacities can be computed as
\begin{align}\label{eq.cb1}
C_B=(1-q)^2+q(1-q)[C(S_1)+1-h(\alpha)]+q^2 C(S_3).
\end{align}
where $\alpha$ is the average bit error rate caused by the avalanche effect. In addition, $C(S_1)$ and $C(S_3)$ are given in Eq. (\ref{eq.cap-13}), implying that these capacities mainly depend on $q$, $\gamma$, $\alpha$ and $\eta$ . As a result, the main channel capacity depends on $q$ and $\gamma$ which according to Eq.'s (\ref{eq.ers1c}) and (\ref{eq.gama}) are themselves functions of $\eta$, for a fixed $\alpha$. Therefore, Bob's channel capacity mainly depends on the original channel cross over probability $\eta$. 

\subsection{Secrecy Capacity of the Wire-tap Channel with Noise}\label{sec14}

As discussed in subsection \ref{sec9}, wiretap channel is a degraded version of the main channel that only in correct key state can be as good as Bob's channel. In the worst possible scenario, we assume she is perfectly aware of channel states. When Eve with the probability of $P_c$ obtains the right key, her channel capacity will be the same as Bob's, i.e. $C_B$, but when with the probability of $P_w$ gets a wrong key, her channel will turn into a BSC with the cross over probability of $\gamma$, which has the capacity of $1-h(\gamma)$. Note that, the erasure state does not contribute to the capacity. Hence, Eve's capacity will be
\begin{align}\label{ce-unknown}
C_E=P_w(1-h(\gamma))+P_c C_B,
\end{align}
where $C_B$ is given in Eq. (\ref{eq.cb1}). In the following theorem
secrecy capacity is found whose proof is given in Appendix~\ref{ap.d}.

\begin{thm}
\label{secrecy capacity}
\textit{The secrecy capacity for the created wire-tap channel with the described channel models for Bob and Eve will be}
\begin{gather}\label{cs-known}
C_s=C_B(1-P_c)-(1-P_e-P_c)(1-h(\gamma)).
\end{gather}
\end{thm}

This result implies that secrecy capacity mainly depends on $P_c$,
$P_e$ and $C_B$. Due to the fact that all $P_c$, $P_e$ and $C_B$ highly
depend on the channel error rate $\eta$, the main parameter that impacts
secrecy capacity of the system is intentional noise. Namely,
if Alice can control the cross over probability of the channel, it is
possible to adjust secrecy rate of the system. Note that Alice applies
secrecy encoding over multiple frames in order to statistically average
over Eve's possible failures in frame level, and also to enable Bob to
do the error correction coding when burst of errors occurs. Basically,
Alice and Bob has to use a well designed wiretap channel encoder, based
on the computed secrecy rate in Eq. (\ref{cs-known}). Notably, the main
issue in this scheme is delay that is imposed on the system by applying
multiple frame encoding that makes this scheme applicable only for delay
tolerant communication.

\section{Numerical Results}\label{sec15}

The main objective of numerical analysis is to evaluate the effect of varying $\eta$ on secrecy rate in order to see if there exists an optimum value for $\eta$ for which secrecy capacity reaches its maximum. In simulations, we assume that Alice by controlling $\eta$ is able to generate a degraded wiretap channel. In addition, we assume that the whole data is divided into equal size frames, each containing as many number of $64$-bit data blocks as four-state MC reaches its steady state, such that for each frame, encryption and decryption key remains constant.

Let us assume that $\theta=2^{48}$ is the maximum number of DES encryptions that Eve can perform to establish an attack on each frame. Because for instance, with a CPU having speed of $2.6$ GHz, it takes for about $30$ hours for her to accomplish these many encryptions. For attack optimization algorithm proposed in section \ref{sec6}, the initial values selected for $n$ is $n_0=20$, maximum possible value for $N_c$ is chosen $N_{cmax}=100$, and the thresholds $T_f$ and $T_m$ are set to $10^{-5}$. Furthermore, we chose $\alpha$ as avalanche effect bit error rate to be $0.5$. To evaluate the effect of noise variation on the performance of the system, we changed $\eta$ from $10^{-4}$ to $0.05$ with $500$ steps of size $10^{-4}$. Moreover, suppose that Eve is able to detect these step size changes on $\eta$ by probing the channel and each time is able to optimize all attack parameters using the parameter optimization algorithm.
We assume that Eve is not allowed to use more than $N_{max}=2^{46}$ number of pairs, and prior to attack on each frame, she obtains the required number of plaintext/ciphertext pairs and mounts her attack.

\begin{figure}
\centering
\includegraphics[width=2.5in]{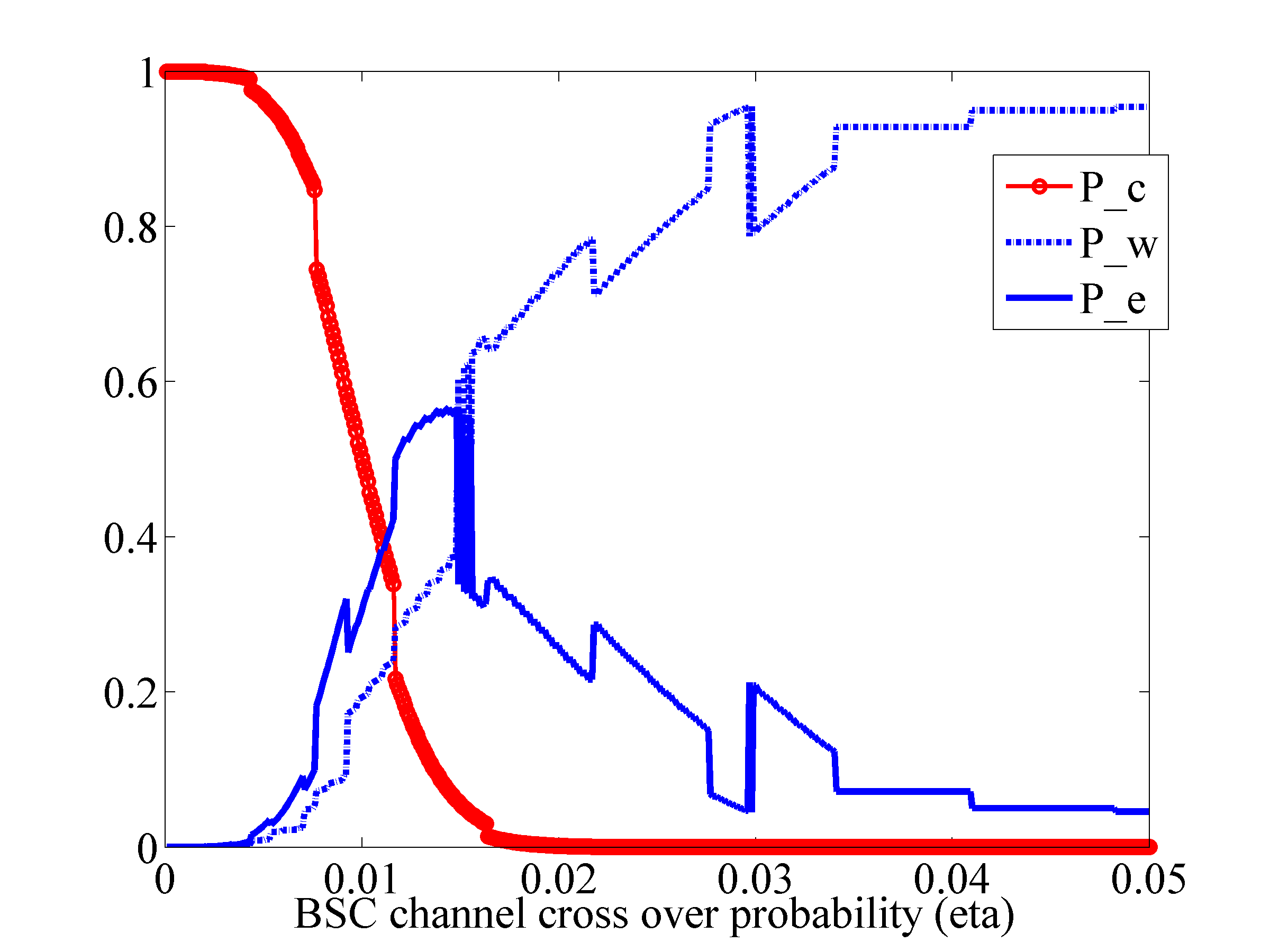}
\caption{Overall success probability, frame erasure and wrong key probabilities versus channel cross over probability} \label{fig11}
\end{figure}

\begin{figure}
\centering
\includegraphics[width=2.5in]{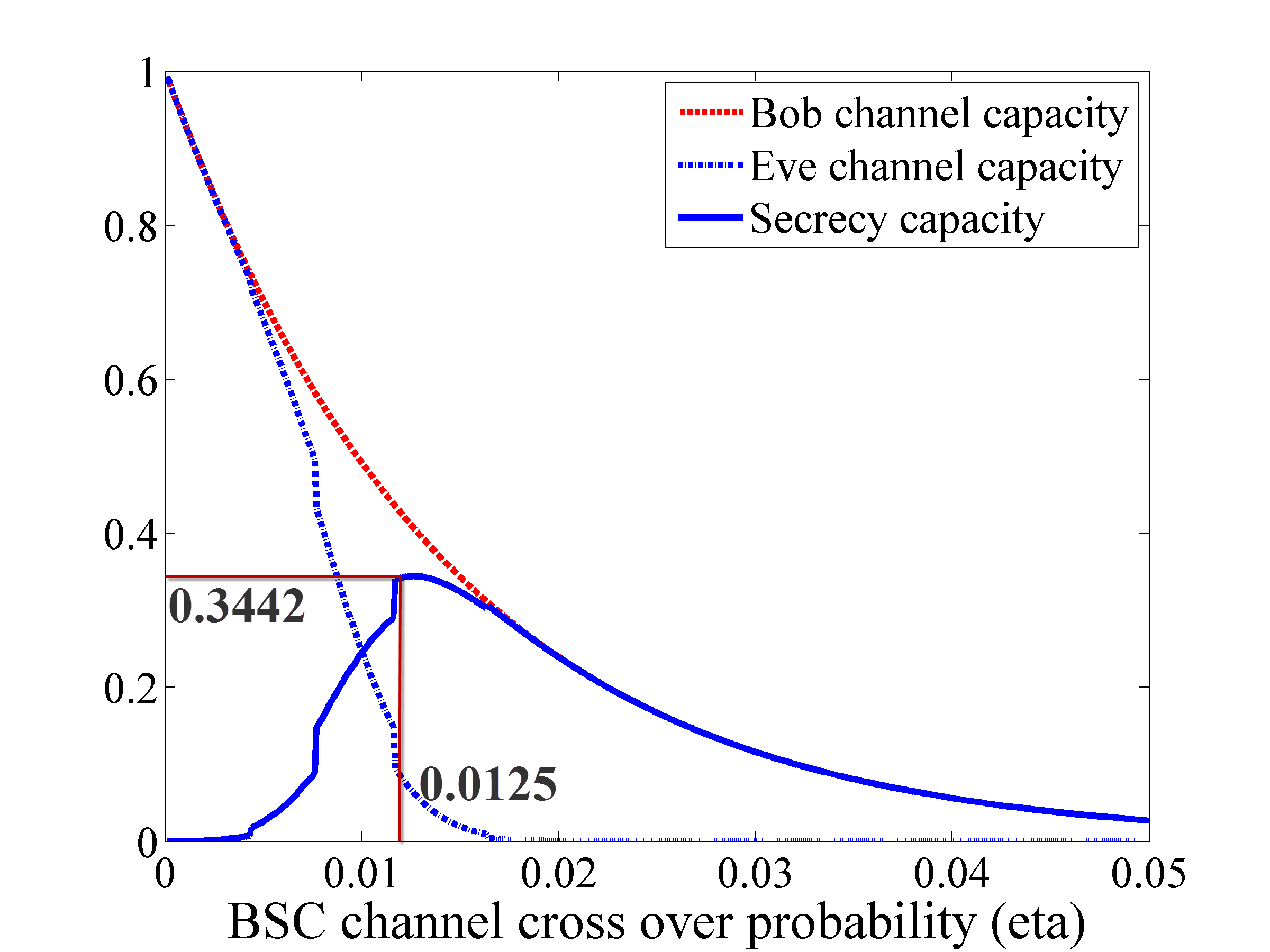}
\caption{Main channel and Eve's channel capacities and secrecy capacity for varying channel cross over probability} \label{fig12}
\end{figure}

In Fig. \ref{fig11}, overall success probability, wrong key and frame erasure probabilities are depicted as functions of $\eta$ for fixed number of pairs equal to $2^{46}$. As this Figure displays with rising $\eta$, $P_c$ is monotonically decreasing, reaching zero for $\eta>0.017$, while wrong key probability $P_w$ goes to $1$ for $\eta=0.05$ because of increase in $P_F$. As discussed in section \ref{sec4}, the obtained results for $P_w$ show that it becomes considerable for some channel conditions and can not be ignored. In Fig. \ref{fig12} curves of main and wiretap channel capacities as well as the secrecy capacity are drawn as functions of $\eta$. This Figure shows that Alice-Bob channel capacity is monotonically decreasing with increase in $\eta$ while secrecy capacity $C_s$ rises up to its maximum value $0.3442$ for $\eta=0.0125$ and then falls. 
Indeed, this cross over probability can be considered optimum value for which secrecy capacity achieves its maximum.
\begin{table}[h!t!p!]
\caption{Optimized attack parameters using proposed algorithm in \ref{sec6}}
\centering
\begin{tabular}{|c|c|c|c|c|}
  \hline
  $\eta$ & $0.001$ & $0.005$ & $0.01$ & $0.0125$ \\
  \hline
  $N_c$ & $5$ & $9$ & $16$ & $20$  \\
  \hline
  $\tau$ & $3$ & $5$ & $6$ & $7$ \\
  \hline
  $a$ & $23$ & $24$ & $24$ & $27$ \\
  \hline
  $P_c$ & $0.9999$ & $0.9636$ & $0.5014$ & $0.1618$ \\
  \hline
  \end{tabular}
  \label{tab2}
\end{table}

In Table \ref{tab2} optimized attack parameters using our proposed algorithm for four different $\eta$'s, i.e. $0.001, 0.005, 0.01$ and $0.0125$ are given. According to this table, with increase in $\eta$, the required number of trials $N_c$ for each key increases from $5$ to $20$ in order to keep $P_m$ below the threshold  $T_m=10^{-5}$ when it rises. The same holds for parameters $a$ and $\tau$ which to achieve the determined thresholds, have to increase with rising channel noise to maximize the overall success probability. According to our numerical results, Alice can adjust channel conditions by introducing deliberate noise in application layer to have $\eta=0.0125$, to achieve the desirable secrecy capacity.

\section{Conclusion}\label{sec16}

In this paper we showed that by introducing tunable noise in
application layer upon the encrypted data, even though Eve utilizes
an optimized attack strategy, the secrecy rate of the system can
remarkably increase. In fact, Alice can achieve a sufficiently large
secrecy capacity by adjusting the cross over probability of the channel
using deliberate noise. This secrecy rate guarantees a highly secure and
reliable communication using wiretap channel coding in application
layer over multiple frames. For secrecy capacity computation we tailored the known channel states scenario.
In our future work, we will focus on the unknown state case and also will consider a more generic cipher. In addition, we
will work on more detailed design of a secrecy encoder in this framework.
\appendices
\section{Proof of Theorem \ref{total_success}}\label{appendix.a}
\begin{proof}
Suppose that all possible $2^{56-a}$ key candidates are arranged as $k_1, k_2, \ldots, k_{2^{56-a}}$ from the lowest rank to the highest. Let $H_i$ be the hypothesis that $k_i$ is the original key and $g_i=1$ be the event that Eve decides that $k_i$ is correct. We define a Bernoulli random variable $B$ which is equal to $1$ when the right key is among top top $2^{56-a}$ candidates, and $0$, otherwise. Thus, $Pr[B=0]=1-P_s$ and $Pr[B=1]=P_s$. Let $P_c$ be the total success probability for Eve. Note that when $B=0$, the right key will not be tested and consequently can not be found. Therefore, we have
\begin{gather}\label{eq.pc1}
P_c=\sum_{i=1}^{2^{56-a}}Pr[g_i=1,H_i|B=1].Pr[B=1].
\end{gather}
The probability that Eve can realize the right key $k_i$ is
\begin{align}
Pr[g_i=1,H_i|B=1]=Pr[g_i=1|H_i,B=1].Pr[H_i|B=1].\nonumber
\end{align}
For Eve to be able to find the correct key at rank $i$, since she starts the test from upper ranks to the lower ones, there should not be any false key acceptance for ranks higher than $i$, as well as a key missing event for rank $i$. Hence,
\begin{gather}\label{eq.suconekey}
Pr[g_i=1|H_i,B=1]=(1-P_F)^{2^{56-a}-i}(1-P_m).
\end{gather}

Moreover, Decisions about all $2^{56-a}$ keys are independent, and all of the tested keys are equally probable to be the right one, i.e. $Pr[H_i|B=1]=\frac{1}{2^{56-a}}$. Therefore, by using Eq.'s (\ref{eq.pc1}) and (\ref{eq.suconekey}), we obtain Eq. (\ref{eq.pc}) for total success probability.

The next step is to compute the frame erasure probability. Assume that the right key is $k_i$ and is located among top $2^{56-a}$ candidates. In order to obtain no key, Eve should not have any false key admission for $k_j$, $j\neq i$ for $i,j=2^{56}-2^{56-a}+1,\ldots,2^{56}$, i.e. top $2^{56-a}$ candidates except the right key itself, and in addition to that she has to miss the right key $k_i$. When $k_i$ is not among top candidates, since it will not be examined, Eve gets nothing provided that there has been no wrong key acceptance event for top $2^{56-a}$ tested candidates. As a result, frame erasure probability can be computed according to Eq. (\ref{eq.pe}). By a similar technique, we can prove that the wrong key probability is $P_w=1-P_e-P_c$.
\end{proof}
\section{Proof of Lemma \ref{Vector transition probability}}\label{ap.B}
\begin{proof}
We need to compute vector transition probabilities between all possible input and output vectors $X$ and $Y$ for states $S_1$ and $S_3$. Hence, for $k=1,3$
\begin{align}\label{eq.ersk}
Pr[Y|X,S_k]&=Pr[X \oplus Y|X,S_k]=Pr[E|S_k],
\end{align}
where $E$ is the decryption error vector which is bit-wise Xor of input and output vectors. The last equality is because $E$ depends on channel errors in previous and current ciphertexts, so given the state, it is independent from input vector $X$. To analyze states $S_0$ and $S_1$, we define two events, $A$ and $B$ as
\begin{align}
A:&\quad \textrm{There exists at least one bit error in}\quad \hat{C}_i\nonumber\\
B:&\quad \textrm{There exists at least one bit error in}\quad \hat{C}_{i-1}.\nonumber
\end{align}
As a result, $S_1=A \cap \bar{B}$, and we can write
\begin{align}\label{eq.ers1b}
Pr(E|S_1)=&Pr(E|A,\bar{B})=\frac{Pr(E|\bar{B})Pr(A|E,\bar{B})}{Pr(A|\bar{B})}.
\end{align}

The fact that events $A$ and $B$ are caused by two independent channel error vectors $Z_{i-1}$ and $Z_i$ implies that $A$ is independent of $B$ and its complementary, i.e. $Pr(A|\bar{B})=Pr(A)=q$. When event $B$ has not occurred, since only $\hat{C_i}$ can induce bit errors with rate of $\eta$ into the stored plaintext, the probability that a particular decryption error vector $E$ with Hamming weight of $W(E)$ takes place will be
\begin{gather}\label{eq.ers1d}
Pr(E|\bar{B})=\eta^{W(E)}(1-\eta)^{64-W(E)}.
\end{gather}

In state $S_1$, HW of error vector $E$ can not be zero because we know that the only source that can induce error at stage $i$ is $Z_i$ that surely has a non-zero bit. In this case, given an error vector $E$ with $W(E)\neq 0$ and knowing that event $B$ did not occur, we can infer that this error is induced by error in $\hat{C}_i$, hence event $A$ has certainly occurred, i.e. $Pr(A|E,\bar{B})=1$. Thus, using equations (\ref{eq.ersk}), (\ref{eq.ers1b}) and (\ref{eq.ers1d}), we can obtain the input-output transition probability in $S_1$ as in Eq. (\ref{eq.ers1}).

On the other hand, according to its definition, state $S_3$ takes place when both events $A$ and $B$ happen, i.e. $S_3=A \cap B$. Therefore,
\begin{gather}\label{eq.ers3a}
Pr(E|S_3)=Pr(E|A,B)=\frac{Pr(E|B)Pr(A|E,B)}{Pr(A|B)}.
\end{gather}

Knowing that $B$ occurred, implies that there exists one bit error in DES input, which induces independent bit errors with the rate of $\alpha$ in cipher output and consequently in $\hat{P}_i$, but also there is independent bit error sequence caused by $\hat{C}_i$ that has the rate of $\eta$. Since the decryption error vector $E_i$ is a result of Xor of these two error sequences, we can say that $E_i$ is a sequence of random bits with i.i.d. distribution and bit error probability of $\gamma$ which is given in Eq. (\ref{eq.gama}). As a result,
\begin{gather}\label{eq.ers3b}
Pr(E|B)=\gamma^{W(E)}(1-\gamma)^{64-W(E)},
\end{gather}
so we can write
\begin{align}\label{eq.ers3c}
Pr(A|E,B)=1-Pr(\bar{A}|E,B)=1-\frac{Pr(E|\bar{A},B)Pr(\bar{A}|B)}{Pr(E|B)}.
\end{align}

When $A$ has not occurred, But $B$ has, the only error source will be the cipher input that induces independent bit errors with the rate of $\alpha$ into the output. Consequently, we have
\begin{gather}\label{eq.ers3d}
Pr(E|\bar{A},B)=\alpha^{W(E)}(1-\alpha)^{64-W(E)}.
\end{gather}

Then, using equation (\ref{eq.ers3b}), (\ref{eq.ers3c}) and (\ref{eq.ers3d}) gives us
\begin{gather}\label{eq.ers3e}
Pr(A|E,B)=1-\frac{\alpha^{W(E)}(1-\alpha)^{64-W(E)}(1-q)}{\gamma^{W(E)}(1-\gamma)^{64-W(E)}}.
\end{gather}

Finally, according to equations (\ref{eq.ersk}), (\ref{eq.ers3a}) and (\ref{eq.ers3e}) the input-output transition probability in state $S_3$ for all $W(E)$ can be obtained using Eq. (\ref{eq.ers3}).
\end{proof}
\section{Proof of Lemma \ref{Entropy}}\label{ap.c}
\begin{proof}If we assume that all $2^{64}$ possible input plaintexts are equally likely, for $l=1,3$ we can write
\begin{align}\label{eq.mut-s1-a}
H(Y|S_l,X)=&-\sum_i\sum_j \frac{1}{2^{64}} Pr(Y=Y_j|X=X_i,S_l)\\
&.\log Pr(Y=Y_j|X=X_i,S_l)\nonumber\\
=&-\sum_{j=1}^{2^{64}} Pr(E=E_j|S_l)\log Pr(E=E_j|S_l).\nonumber
\end{align}
The second equality is resulted from Eq. (\ref{eq.ersk}) for $E_{i,j}=X_i\oplus Y_j$ as the decryption error vector. Furthermore, for state $S_1$ as discussed in subsection \ref{sec8}, Hamming weight of the error vector $E$ can not be zero. Thus, we can take $E_1$ as a $64$-bit zero vector and exclude it from this summation. Then, using Eq. (\ref{eq.ers1}) brings about the following result 
\begin{align}\label{eq.mut-s1-c}
H(Y|S_1,X)=&\frac{-1}{q}\sum_{j=2}^{2^{64}}\eta^{W(E_j)}(1-\eta)^{64-W(E_j)}\nonumber\\
&.\log\left[\frac{\eta^{W(E_j)}(1-\eta)^{64-W(E_j)}}{q}\right].
\end{align}
We know that out of all $2^{64}$ error vectors, the number of possible vectors with Hamming weight of $W$ or vectors with $W$ non-zero bits is the number of possibilities of choosing $W$ bits out of $64$ bits which is equal to $W$-combinations from $64$ elements. Finally, Eq. (\ref{eq.mut-s1-c}) can be rewritten as Eq. (\ref{eq-ent-s1}). Note that we excluded zero weight case, i.e. $k=0$.

For state $S_3$, we can compute $H(Y|S_3,X)$ using Eq. (\ref{eq.mut-s1-a}) for $l=3$. In this case, $j=1$ is not excluded because unlike state $S_1$ in state $S_3$, it is possible to have decryption error vector $E_1$ with zero weight. Finally, by using Eq. (\ref{eq.ers3}) similar to the entropy in $S_1$, we obtain $H(Y|S_3,X)$ in Eq. (\ref{eq-ent-s3}).
\end{proof}
\section{Proof of Theorem \ref{secrecy capacity}}\label{ap.d}
\begin{proof} It is shown in \cite{Dijk} that when the mutual information between Alice at Bob and the mutual information between Alice and Eve are individually maximized by the the same input distribution, and the main channel is less noisy that the wiretap channel, the secrecy capacity can be computed as the difference of two capacities. In our channel model, since both Bob and Eve's mutual information with Alice are maximized with uniformly distributed inputs $X$, and wiretap channel is noisier that the main channel, the secrecy capacity will be, $C_s=C_B-C_E$. It gives us the final result in Eq. (\ref{cs-known}).
\end{proof}

\bibliographystyle{IEEEtran}
\bibliography{journal1}

\begin{thebibliography}{10}
\providecommand{\url}[1]{#1}
\csname url@samestyle\endcsname
\providecommand{\newblock}{\relax}
\providecommand{\bibinfo}[2]{#2}
\providecommand{\BIBentrySTDinterwordspacing}{\spaceskip=0pt\relax}
\providecommand{\BIBentryALTinterwordstretchfactor}{4}
\providecommand{\BIBentryALTinterwordspacing}{\spaceskip=\fontdimen2\font plus
\BIBentryALTinterwordstretchfactor\fontdimen3\font minus
  \fontdimen4\font\relax}
\providecommand{\BIBforeignlanguage}[2]{{%
\expandafter\ifx\csname l@#1\endcsname\relax
\typeout{** WARNING: IEEEtran.bst: No hyphenation pattern has been}%
\typeout{** loaded for the language `#1'. Using the pattern for}%
\typeout{** the default language instead.}%
\else
\language=\csname l@#1\endcsname
\fi
#2}}
\providecommand{\BIBdecl}{\relax}
\BIBdecl

\bibitem{Ehrsam}
W.~F. Ehrsam, S.~M. Matyas, C.~H. Meyer, and W.~L. Tuchman, ``A cryptographic
  key management scheme for implementing the data encryption standard,''
  \emph{IBM Systems Journal}, vol.~17, no.~2, pp. 106--125, 1978.

\bibitem{Schneier:1995}
B.~Schneier, \emph{Applied cryptography (2nd ed.): protocols, algorithms, and
  source code in C}.\hskip 1em plus 0.5em minus 0.4em\relax New York, NY, USA:
  John Wiley \& Sons, Inc., 1995.

\bibitem{Wyner}
A.~Wyner, ``The wire-tap channel,'' \emph{Bell Syst. Tech. J.}, vol.~54, pp.
  1355--1387, 1975.

\bibitem{Csiszar}
I.~Csiszar and J.~Korner, ``Broadcast channels with confidential messages,''
  \emph{IEEE Trans. Inform. Theory}, p. 339–348, May 1978.

\bibitem{Thangaraj}
A.~Thangaraj, S.~Dihidar, A.~Calderbank, S.~McLaughlin, and J.~Merolla,
  ``Applications of {LDPC} codes to the wiretap channel,'' \emph{IEEE
  Transactions on Information Theory}, vol.~53, no.~8, pp. 2933 --2945, Aug
  2007.

\bibitem{Yin}
R.~Yin, S.~Wei, J.~Yuan, X.~Shan, and X.~Wang, ``Tradeoff between reliability
  and security in block ciphering systems with physical channel errors,''
  \emph{Proc. IEEE Military Commun. Conf. (MILCOM)}, 2010.

\bibitem{Goel}
S.~Goel and R.~Negi, ``Guaranteeing secrecy using artificial noise,''
  \emph{Wireless Communications, IEEE Transactions on}, vol.~7, no.~6, pp. 2180
  --2189, june 2008.

\bibitem{Vilela}
J.~Vilela, M.~Bloch, J.~Barros, and S.~McLaughlin, ``Wireless secrecy regions
  with friendly jamming,'' \emph{Information Forensics and Security, IEEE
  Transactions on}, vol.~6, no.~2, pp. 256 --266, june 2011.

\bibitem{Mihaljevic}
M.~J. Mihaljevic and H.~Imai, ``An approach for stream ciphers design based on
  joint computing over random and secret data,'' \emph{Computing}, vol.~85, pp.
  153--168, 2009.

\bibitem{Willett}
M.~Willett, ``Deliberate noise in a modern cryptographic system (corresp.),''
  \emph{IEEE Transactions on Information Theory, vol.26, no.1}, pp. 102-- 104,
  1980.

\bibitem{Mihal}
M.~Mihaljevic� and F.~Oggier, ``A wire-tap approach to enhance security in
  communication systems using the encoding-encryption paradigm,'' \emph{IEEE
  17th International Conference on Telecommunications (ICT)}, pp. 83--88, April
  2010.

\bibitem{Kocher:1999:DPA}
P.~C. Kocher, J.~Jaffe, and B.~Jun, ``Differential power analysis,'' in
  \emph{Proceedings of the 19th Annual International Cryptology Conference on
  Advances in Cryptology}, ser. CRYPTO '99.\hskip 1em plus 0.5em minus
  0.4em\relax London, UK: Springer-Verlag, 1999, pp. 388--397.

\bibitem{Roche}
T.~Roche, V.~Lomné, and K.~Khalfallah, ``Combined fault and side-channel
  attack on protected implementations of {AES},'' \emph{CARDIS}, pp. 65--83,
  2011.

\bibitem{Yijun}
Y.~Liu, P.~Chen, G.~Xie, Z.~Liu, and Z.~Li, ``The design of a low-power
  asynchronous {DES} coprocessor for sensor network encryption,'' in
  \emph{International Symposium on Computer Science and Computational
  Technology (ISCSCT)}, vol.~2, Dec 2008, pp. 190--193.

\bibitem{Zibideh}
W.~Zibideh and M.~Matalgah, ``Modified-{DES} encryption algorithm with improved
  {BER} performance in wireless communication,'' in \emph{Radio and Wireless
  Symposium (RWS), 2011 IEEE}, jan. 2011, pp. 219 --222.

\bibitem{Xiao}
Y.~Xiao, H.~Chen, X.~Du, and M.~Guizani, ``Stream-based cipher feedback mode in
  wireless error channel,'' \emph{IEEE Trans. Wireless Comm.}, vol.~8, pp.
  622--626, 2009.

\bibitem{Heys1}
H.~Heys and S.~Tavares, ``Avalanche characteristics of substitution-
  permutation encryption networks,'' \emph{IEEE Trans. Comput.}, vol. 44, no.
  9, pp. 1131--1139, Sep 1995.

\bibitem{Nyberg}
K.~Nyberg, ``S-boxes and round functions with controllable linearity and
  differential uniformity,'' \emph{in FSE}, pp. 111--130, 1994.

\bibitem{Matsui1}
M.~Matsui, ``Linear cryptanalysis method for {DES} cipher,'' \emph{Lecture
  Notes in Computer Science}, vol. 765, pp. 385--397, 1994.

\bibitem{Selcuk}
A.~Selcuk and A.~Bicak, ``On probability of success in linear and differential
  cryptanalysis,'' \emph{SCN 2002}, pp. 174--185, 2003.

\bibitem{Bennett}
C.~H. Bennett, G.~Brassard, C.~Crpeau, and U.~M. Maurer, ``Generalized privacy
  amplification,'' \emph{IEEE Trans. Inform. Theory, vol. 41}, pp. 1915--1923,
  Nov. 1995.

\bibitem{DESEXP}
M.~Matsui, ``The first experimental cryptanalysis of the data encryption
  standard,'' \emph{Lecture Notes in Computer Science}, vol. 835, pp. 1--11,
  1994.

\bibitem{Ross}
S.~Ross, ``Introduction to probability models,'' \emph{University of Southern
  California, Academic Press}, Tenth Edition, ISBN: 978-0-12-375686-2, 2010.

\bibitem{Yu}
S.~Yu, Z.~Liu, M.~Squillante, C.~Xia, and L.~Zhang, ``A hidden semi-$m$arkov
  model for web workload self-similarity,'' \emph{21st IEEE International
  Performance, Computing, and Communications Conference}, pp. 65--72, 2002.

\bibitem{Goldsmith}
A.~J. Goldsmith and P.~P. Varaiya, ``Capacity, mutual information, and coding
  for finite-state markov channels,'' \emph{IEEE Trans. Inform. Theory},
  vol.~43, pp. 868--886, May 1996.

\bibitem{Holliday}
T.~Holliday, A.~Goldsmith, and P.~Glynn, ``Capacity of finite state markov
  channels with general inputs,'' \emph{In Proceedings of the IEEE
  International Symposium on Information Theory 289}, 2003.

\bibitem{Wang}
H.~S. Wang and N.~Moayeri, ``Finite-state markov channel: A useful model for
  radio communication channel,'' \emph{Proc. IEEE Veh. Tech. Conf. (VTC)},
  vol.~44, pp. 163--171, Feb 1995.

\bibitem{Mushkin}
M.~Mushkin and I.~Bar-David, ``Capacity and coding for the {G}ilbert {E}lliot
  channel,'' \emph{IEEE Trans. Inform. Theory}, vol.~35, pp. 1277--1290, 1989.

\bibitem{Lapidoth}
A.~Lapidoth and I.~E. Telatar, ``The compound channel capacity of a class of
  finite-state channels,'' \emph{IEEE Trans. Inform. Theory}, vol.~44, pp.
  973--983, 1998.

\bibitem{yogesh}
Y.~Sankarasubramaniam, A.~Thangaraj, and K.~Viswanathan, ``Finite-state wiretap
  channels: Secrecy under memory constraints,'' \emph{Information Theory
  Workshop, 2009. ITW 2009. IEEE}, pp. 115 --119, Oct. 2009.

\bibitem{Ozarow}
L.~H. Ozarow and A.~D. Wyner, ``Wire-tap channel {II},'' \emph{Bell System
  Technical Journal}, vol.~63, no.~10, pp. 2135--2157, Dec 1984.

\bibitem{Dijk}
M.~v.~Dijk, ``On a special class of broadcast channels with confidential
  messages,'' \emph{IEEE Trans. Inform. Theory}, vol.~43, pp. 712--714, Mar
  1997.

\end{thebibliography}

\end{document}